\documentclass[pra,showpacs,10pt,preprint,superscriptaddress,floatfix,showkeys]{revtex4-1}
\usepackage{graphicx}
\usepackage{amsmath}
\usepackage{epstopdf}
\usepackage[usenames,dvipsnames,svgnames,table]{xcolor}
\usepackage{upgreek}

\begin{document}

\newcommand{\HaraldComment}[1]{\textcolor{red}{\textit{#1}}}
\newcommand{\JimComment}[1]{\textcolor{orange}{\textit{#1}}}
\newcommand{\RenateComment}[1]{\textcolor{blue}{\textit{#1}}}

\newcommand{\degree}{^\circ}
\newcommand{\unit}[1]{\,\mathrm{#1}}
\newcommand{\Figureref}[2][]{Figure~\ref{#2}#1}
\newcommand{\Eqnref}[1]{Equation~(\ref{#1})}
\newcommand{\bra}[1]{\left<\mathrm{#1}\right|}
\newcommand{\ket}[1]{\left|\mathrm{#1}\right>}

\newcommand{\affilOU}{Homer L. Dodge Department of Physics and Astronomy, The University of Oklahoma, 440 W. Brooks St. Norman, OK 73019, USA}
\newcommand{\affilStutt}{5. Physikalisches Institut, Universit\"{a}t Stuttgart, Pfaffenwaldring 57 D-70550 Stuttgart, Germany}

\title{Atom-Based Sensing of Weak Radio Frequency Electric Fields Using Homodyne Readout}
\author{Santosh Kumar}
\affiliation{\affilOU}
\author{Haoquan Fan}
\affiliation{\affilOU}
\author{Harald K\"{u}bler}
\affiliation{\affilStutt}
\author{Jiteng Sheng}
\affiliation{\affilOU}
\author{James P. Shaffer}
\affiliation{\affilOU}
\email{Corresponding author: shaffer@nhn.ou.edu}

\date{\today}
\begin{abstract}
{We utilize a homodyne detection technique to achieve a new sensitivity limit for atom-based, absolute radio-frequency electric field sensing of $\mathrm{5 \mu V cm^{-1} Hz^{-1/2} }$. A Mach-Zehnder interferometer is used for the homodyne detection. With the increased sensitivity, we investigate the dominant dephasing mechanisms that affect the performance of the sensor. In particular, we present data on power broadening, collisional broadening and transit time broadening. Our results are compared to density matrix calculations. We show that photon shot noise in the signal readout is currently a limiting factor. We suggest that new approaches with superior readout with respect to photon shot noise are needed to increase the sensitivity further.}
\end{abstract}

\maketitle

\section{Introduction}

Atom-based measurements have been successfully utilized for magnetometery \cite{koschorreck2010sub,wasilewski2010quantum,balabas2010polarized,savukov2005tunable}, time and frequency standards \cite{ludlow2015optical}, inertial force sensing \cite{cronin2009optics} as well as searches for local Lorentz invariance \cite{chung2009atom,smiciklas2011new,bear2000limit} and intrinsic electric dipole moments of the neutron \cite{baker2006improved} and electron \cite{baron2014order}, amongst others. The accuracy and repeatability of atom-based measurements significantly surpass conventional methods because the stable properties of atoms and molecules are advantageous for precision measurement.  Recently, Rydberg atoms have been introduced to measure the amplitude of radio frequency (RF) electric fields following the same rationale \cite{sedlacek2012microwave}. For Rydberg atom-based RF electric field sensing, electromagnetically induced transparency (EIT) is used to readout the effect of a RF electric field on atoms contained in a vapor cell at room temperature \cite{fan2015atom,Hollowayrev}. The possibility of performing high resolution Rydberg atom spectroscopy in micron sized vapor cells is an important enabler of the method \cite{kubler2010coherent,fan2015effect}, particularly at higher frequencies. The Rydberg atom-based RF electric field measurement is promising for performing traceable measurements with a higher sensitivity, accuracy and stability than conventional antenna-based standards. Consequently, Rydberg atom-based RF electrometry has widespread applications in areas such as antenna calibration, signal detection, terahertz sensing and the characterization of electronics and materials in the RF spectrum.

The current sensitivity of Rydberg atom-based RF electric field sensing is $\mathrm{\sim30~\mu V cm^{-1}Hz^{-1/2}}$ \cite{sedlacek2012microwave}. Imaging \cite{fan2014subwavelength,gordon2014millimeter,Adamsterahertz} and vector detection \cite{sedlacek2013atom} are possible. The high sensitivity of Rydberg atom-based RF electric field measurement is the result of the large transition dipole moments between Rydberg states, $\sim 100-10000\,$e$\,$a$_0$ depending on the transition \cite{gallagher2005rydberg}. The readout method effectively prepares each participating atom as an interferometer so that the RF electric field induces changes in the light-matter interaction that can be detected optically. The shot noise, or projection noise, limited sensitivity of a collection of atoms in a vapor cell is several orders of magnitude higher, $\sim 4$, than what has been realized so far, depending on the frequency and other parameters, such as vapor cell gas density \cite{fan2015atom}.

Noise in the readout of the signal due to the EIT probe laser can be a limiting factor for the sensitivity, as well as fundamental processes such as photon shot noise on the associated detector. The probe laser noise is due to changes in power, frequency and polarization. In cases where the predominant noise is random, it is often prudent to increase the power of the probe laser to increase the signal-to-noise-ratio (SNR). For Rydberg atom-based RF electric field sensing, it is not possible to simply turn up the probe laser power for several reasons. To effectively use the large transition dipole moments of Rydberg atoms, the oscillations of the Rydberg transition dipole must be coherent. The population of Rydberg atoms and ground state atoms has to be low enough inside the vapor cell to reduce collision rates so that the coherence time of the atoms is sufficiently long to achieve the target sensitivity. The long ranged interactions between Rydberg atoms yield large collision and ionization cross-sections \cite{AAMOP2014}. Photoionization and blackbody radiation can also become problematic. Collision rates between Rydberg atoms and ground state atoms can be reduced by using lower vapor pressures but the desire to realize a spectrally narrow EIT feature also pushes the measurements towards low probe Rabi frequencies. The sensitivity improves as more atoms participate, but at the same rate that loss of coherence time degrades the sensitivity as both collision rates and atom number are proportional to the atom density. Negotiating these factors restricts the Rabi frequencies used for the measurement.

An optical interferometer is one option for reducing noise from the probe laser. Many experiments with atoms \cite{denschlag2000generating,pezze2005sub}, photons \cite{purves2006sagnac,rarity1990two,xiao1995measurement,starling2010continuous,zibrov1996experimental}, and electrons \cite{ji2003electronic} show that interferometers have the ability to perform high sensitivity measurements with the potential to reach the shot noise limit \cite{armen2002adaptive}. We use a Mach-Zehnder interferometer (MZI) along with a homodyne detection technique \cite{oliver2005mach,rarity1990two,xiao1995measurement,aljunid2009phase,scully1992high,armen2002adaptive} to improve the measurement sensitivity of Rydberg atom-based RF electric field sensing. MZIs have been widely used in various fields for precision measurement to achieve shot noise limited sensitivities \cite{scully1992high,louchet2010entanglement,lan2012influence,trubko2015atom,cronin2009optics}. In this paper, we used a free-space interferometer as proof of principle to approach photon shot noise limited performance in Rydberg atom-based RF electric field sensing. Fiber or chip based MZIs can be implemented for a compact RF electric field sensor \cite{Ritter_APL2015}.

The MZI detects the nonlinear phase shift instead of directly measuring the transmitted probe power, in contrast to our prior work \cite{sedlacek2012microwave,sedlacek2013atom,fan2015effect,dispersive}. The noise of the probe laser is reduced by the subtraction taking place in the homodyne detection and the EIT signal is enhanced by the strong local oscillator (LO). We achieved a sensitivity of $\mathrm{\sim 5\mu V cm^{-1} Hz^{-1/2}}$, which is six times better than our previously reported result \cite{sedlacek2012microwave}. The increased SNR provides an opportunity to quantitatively study factors needed to optimize the sensitivity. We study power broadening, collision broadening and transit time broadening.




\section{Materials and Methods}
We use the Cs $6S_{1/2} (F=4)\leftrightarrow 6P_{3/2} (F'=5)\leftrightarrow52 D_{5/2}$ EIT system. The probe transition is the Cs $6S_{1/2} (F=4)\leftrightarrow 6P_{3/2} (F'=5)$  transition while the coupling transition is the Cs $6P_{3/2} (F'=5)\leftrightarrow52 D_{5/2}$ transition. The RF electric field is tuned to resonance with the $52 D_{5/2}\leftrightarrow 53 P_{3/2}$ Rydberg transition.

\begin{figure}[htbp]
\centering
\includegraphics[width=\columnwidth]{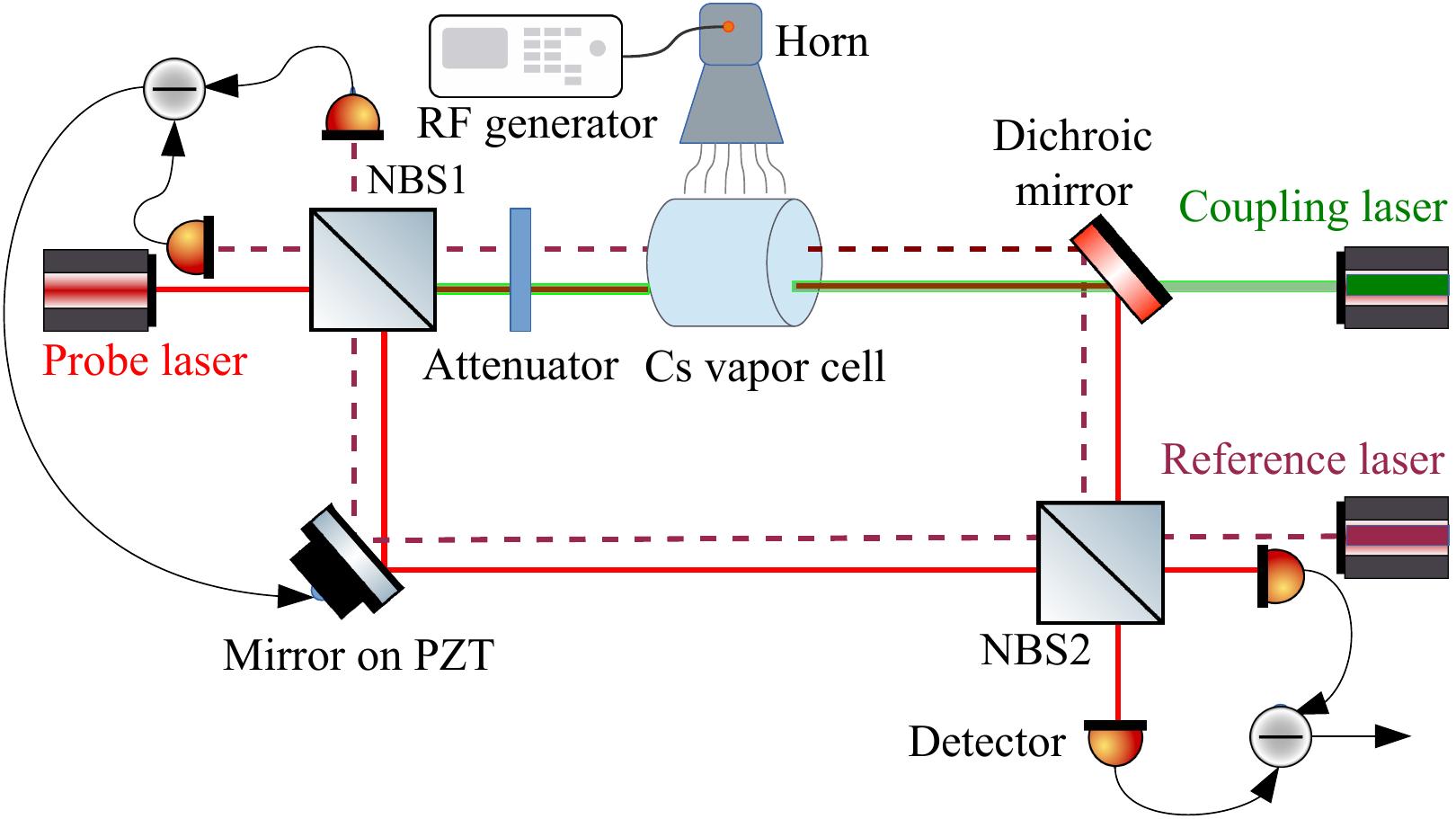}
\caption{This figure shows a schematic of the experimental setup. The probe laser beam is split into two paths in the MZI by a $50-50$ NBS, NBS1. In one arm of the MZI, referred to as the signal arm,  the probe laser beam passes through a Cs vapor cell. The coupling laser beam counterpropagates along the probe laser beam in the signal arm. The other arm serves as a local oscillator. The probe laser light from the two paths is overlapped in NBS2. The probe laser output from the two ports of NBS2, also a $50-50$ NBS, is captured using a homodyne detection technique. A reference laser enters the MZI from the detection side of the setup and is overlapped with the probe laser in the MZI. The output of the reference laser is detected by a differential photodiode and is sent to a feedback loop to stabilize the MZI. A mirror mounted on a PZT is used to adjust the cavity length to stabilize the relative phase between the beams in the different arms of the MZI.\label{miz_setup}}
\end{figure}

Fig.~\ref{miz_setup} shows the experimental setup. A tunable diode laser is offset locked to an ultrastable Fabry-Perot cavity that is near resonant with the Cs $6S_{1/2} (F=4)\rightarrow 6P_{3/2} (F'=5)$ transition at $\sim 852\,$nm. A $4\,$cm long vapor cell filled with Cs is located in the signal arm of the MZI. The ratio of LO to signal is $\sim 20$. The probe light in the signal and LO arms are recombined at a nonpolarizing beam splitting cube (NBS), NBS2 in Fig.~\ref{miz_setup}, after being split at NBS1. The light in the two arms has the same polarization. The two output channels of the MZI are captured by a pair of photodetectors and the difference signal is measured. We estimate the probe laser linewidth to be $\sim 50\,$kHz based on the locking error signal. The probe laser beam has a nominal size of $1.36 \pm 0.01\,$mm unless otherwise stated.

The coupling laser at $\sim 509\,$nm, resonant with the Cs $6P_{3/2} (F'=5)\leftrightarrow52 D_{5/2}$ Rydberg transition, passes through the signal arm turning mirror and overlaps with the probe laser in a counterpropagating geometry. The coupling laser is also offset locked to an ultrastable Fabry-Perot cavity. The coupling laser is intensity-modulated using an acoustic-optical modulator. The difference signal detected at NBS2 is demodulated with a lock-in amplifier. We estimate the coupling laser linewidth to be $\sim 50\,$kHz based on the locking error signal. The coupling beam has a nominal size of $0.12 \pm 0.01 \,$mm unless otherwise specified.

A horn antenna radiates a RF electric field at a frequency of $5.047\,$GHz to resonantly couple the Cs $ 52 D_{5/2}\leftrightarrow 53 P_{3/2}$ Rydberg states. The vapor cell is placed so that it can be uniformly illuminated. RF absorbing material is placed around the setup to minimize reflections.

A reference laser at $\lambda_r= 795\,$nm is used to lock the phase of the interferometer. The reference laser is locked to a Rb saturated absorption setup. We estimate its linewidth to be $\sim 300\,$kHz. The stability of the MZI is estimated to be $\Delta s\sim 0.4\,$nm, or $\Delta s/2 \pi \times \lambda_r = 8 \times 10^{-5}$. The reference laser is overlapped with the probe beam in the MZI. The output of the reference laser is detected by a pair of photodetectors as shown in Fig.~\ref{miz_setup}. The difference signal is detected and used in a feedback loop to stabilize the MZI.  A piezo-electric transducer (PZT) is used to adjust the path length of the MZI \cite{Locking}.

\begin{figure}[htbp]
\centering
\includegraphics[width=\columnwidth]{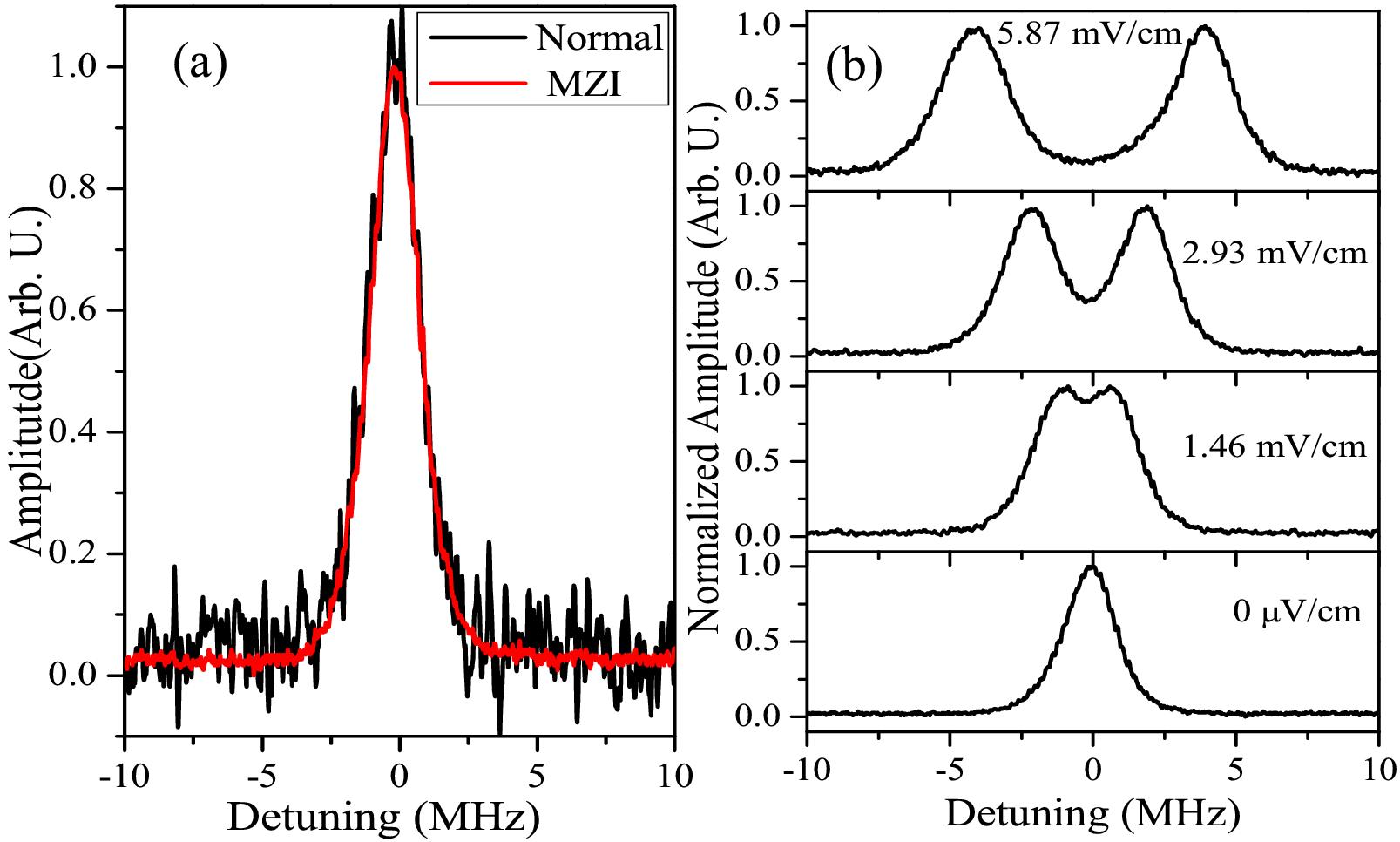}
\caption{(a) Shows a single trace of the Rydberg EIT probe transmission signal with and without MZI. The data is taken using the same Rabi frequencies for both probe and coupling laser which were $2 \pi \times 1.8\pm 0.1\,$MHz and $2 \pi \times 0.50 \pm 0.02\,$MHz respectively. The plot shows the signal as a function of probe laser detuning. The experimental parameters for each signal are the same so the curves are normalized so that the SNR can be compared. (b) Shows single experimental traces demonstrating Autler-Townes splitting using the MZI for different RF electric field amplitudes for the same probe and coupling laser Rabi frequencies as in (a). The scan period is $\mathrm{1~sec}$ and the integration time is $\mathrm{1~ms}$. The signal is plotted as a function of probe laser detuning.\label{MZI_pwscan}}
\end{figure}

To perform the experiments where the temperature was varied, a Polymethylpentene (TPX) oven was built to better control the temperature of the vapor cell. The size of the oven is large enough to place the vapor cell and a small heater inside. The vapor cell and oven have some effect on the incident RF electric field \cite{fan2015effect}, however, in this paper, we focus on measuring the RF electric field at the point where the probe and coupling lasers are overlapped, the interaction region.


Density matrix calculations are carried out to compare the experimental results to theory. The density matrix calculations take into account the three levels of the EIT system, Cs $6S_{1/2} (F=4)\leftrightarrow 6P_{3/2} (F'=5)\leftrightarrow52 D_{5/2}$, and the fourth level that is coupled to the EIT system via the RF electric field,  $ 52 D_{5/2}\leftrightarrow 53 P_{3/2}$. Details of similar calculations can be found in Refs.~\cite{sedlacek2012microwave,sedlacek2013atom,dispersive}.

The time evolution of the density matrix operator, in the presence of decay, is obtained from the Liouville equation,
\begin{equation}
\frac{d}{dt} {\mbox{\boldmath $\rho$}} = -\frac{i}{\hbar} [\textbf{H}, \mbox{\boldmath $\rho$}] + \textbf{L} \mbox{\boldmath $\rho$}, \label{OpticalBlochEq}
\end{equation}
where $\textbf{L}$ is the relaxation matrix and $\textbf{H}$ is the total Hamiltonian \cite{fleischhauer2005electromagnetically}. The sources of relaxation in our system are spontaneous emission of the intermediate state, $\mathrm{6P_{3/2}}$, $\mathrm{\Gamma_0 = 2\pi \times 5.2~MHz}$, and Rydberg state spontaneous decay including blackbody radiation for $\mathrm{52D_{5/2}}$, $\mathrm{\Gamma_{1} = 2\pi \times 3.4~kHz}$ and, $\mathrm{\Gamma_{2} = 2\pi \times 1.6~kHz}$ for $\mathrm{53P_{3/2}}$ \cite{beterov2009quasiclassical}. The parameters that depend on the experimental conditions are also considered in the simulation, which include transit time broadening, $\mathrm{\Gamma_{t}}$, Rydberg-ground state atom collisional dephasing and loss, $\mathrm{\Gamma_{col}}$, laser dephasing, $\mathrm{\Gamma}_{l}$, Rydberg atom-Rydberg atom dephasing and loss, $\mathrm{\Gamma_{Ryd-Ryd}}$, and magnetic dephasing, $\mathrm{\Gamma_{m}}$. The calculations are Doppler averaged to compare to the data.

\section{Results and Discussion}

Fig.~\ref{MZI_pwscan}a shows a comparison of the EIT probe transmission spectra with and without the MZI. Both measurements were carried out with the same experimental parameters at room temperature. The probe Rabi frequency was $\Omega_p = 2 \pi \times 1.8\pm 0.1\,$MHz while the coupling Rabi frequency was $\Omega_c = 2 \pi \times 0.50\pm 0.02\,$MHz. As can be clearly seen from inspection of Fig.~\ref{MZI_pwscan}a, the SNR is substantially improved by using the MZI. The enhancement of the SNR is $\sim 20$.

When the RF electric field is at the $\mathrm{mV cm^{-1}}$ level, Autler-Townes (AT) splitting of the probe transmission spectrum due to the RF electric field can be resolved. The amplitude of the RF electric field can be determined directly by observing a single trace of the probe transmission spectrum because the AT splitting is proportional to the RF electric field amplitude \cite{sedlacek2012microwave}, $\Delta\nu_{AT} = \mu E_{RF}/h$ where $\mu$ is the transition dipole moment and $E_{RF}$ is the RF electric field amplitude and we have assumed that the dipole moment and electric field are parallel and ignored Doppler effects. Fig.~\ref{MZI_pwscan}b shows probe transmission spectra recorded under conditions where the RF electric field causes AT splitting for several different RF electric field amplitudes. The SNR obtained with the MZI shown in Fig.~\ref{MZI_pwscan}b demonstrates that, for these types of RF electric field amplitudes, the RF electric field can be measured in $\lesssim 1\,$s. By increasing the SNR, the MZI enables the measurement to be carried out more quickly, which is important if the RF electric field needs to be determined over a large area, as might be the case for antenna calibration. The increased SNR achieved with the MZI is also promising for measurements where the vapor cell dimensions must be $\ll \lambda_\mathrm{RF}$, where $\lambda_\mathrm{RF}$ is the wavelength of the RF radiation, to avoid perturbing the RF electric field \cite{fan2015effect,fan2015atom}.

\begin{figure}[htbp]
\centering
\includegraphics[scale=0.36]{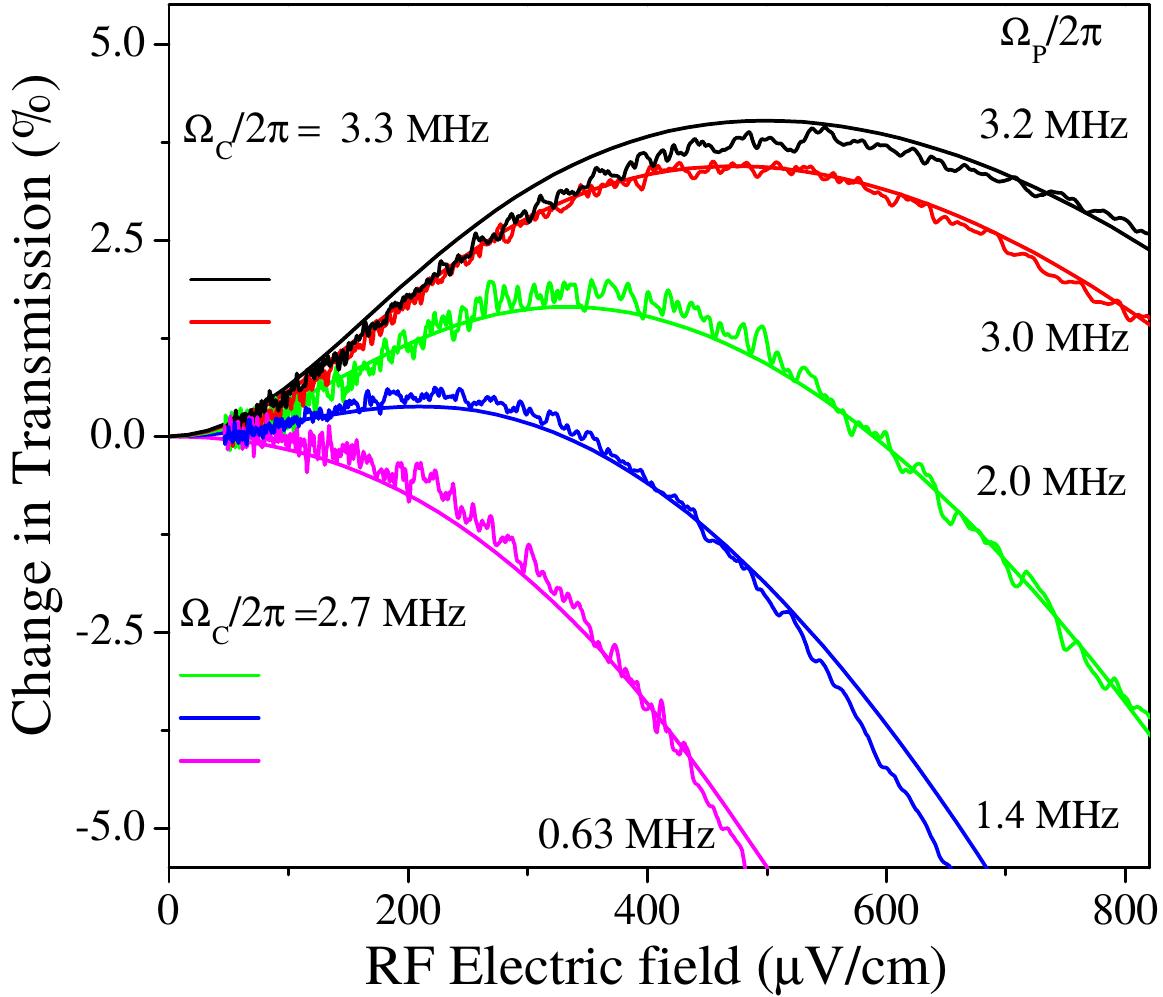}
\caption{Percentage change in transmission vs RF electric field at different probe and coupling laser Rabi frequencies on resonance at room temperature. The black and red curves correspond to $\Omega_c= 2 \pi \times 3.3 \pm 0.1\,$MHz while the green, blue, and magenta curves correspond to $\Omega_c=2 \pi \times 2.7 \pm 0.1\,$MHz. The negative values indicate that the RF electric field reduces the transmission of the probe laser while the positive values mean that the probe laser transmission has been enhanced. The solid curves are four-level density matrix calculations with the corresponding laser and RF Rabi frequencies. The calculations used $\Gamma_l = 2 \pi \times 70\,$kHz, $\Gamma_\mathrm{t}= 2 \pi \times 300\,$kHz, $\Gamma_\mathrm{col} = 2 \pi \times 6\,$kHz, and $\Gamma_\mathrm{m}= 2 \pi \times 50\,$kHz. $\Gamma_\mathrm{Ryd-Ryd} = 2 \pi \times 0\,$kHz because the Rydberg density was kept low in the experiment so it is negligible compared to the other dephasing mechanisms for these experiments. The theoretical plots in the graph have no adjustable parameters.\label{fig:power}}
\end{figure}

When the RF electric field is $\lesssim 1\,\mathrm{mV cm^{-1}}$ the AT splitting is difficult to resolve and the RF electric field amplitude can be determined by measuring the amount of probe laser transmission relative to the probe laser transmission in the absence of the RF electric field on resonance. The residual Doppler shift due to the wavelength mismatch between the coupling and probe lasers limits the spectral resolution, thus determining the conditions at which the AT splitting becomes unresolvable. Making an accurate measurement of the peak transmission amplitude change due to the RF electric field is more difficult than measuring the AT splitting because the effect of the weak RF electric field on the bare three-level EIT lineshape has to be determined. As a consequence, the lineshape must be measured precisely and SNR becomes a larger issue. The MZI is clearly advantageous for such measurements.

It is clear that higher spectral resolution would enable the AT splitting to be observed at lower RF electric field amplitudes. The high SNR demonstrated in Fig.~\ref{MZI_pwscan}a facilitates the detection of the probe transmission at lower probe and coupling laser power compared to the case without the MZI. Thus, power broadening can be reduced in the experimental setup which also can degrade the sensitivity in the AT regime. For example, the full-width-half-maximum (FWHM) of the probe laser EIT spectral bandwidth for the EIT system described in this work is $\sim$1.7 MHz at room temperature. The FWHW of the spectrum of the probe transmission window is smaller than that of previous setups \cite{sedlacek2012microwave} because in our prior work we used higher laser powers for detection to optimize SNR. In a similarly motivated effort, we are currently working on a three-photon scheme for the optical readout that we have proposed for Cs, $6S_{1/2}\leftrightarrow 6P_{1/2} \leftrightarrow 9S_{1/2} \leftrightarrow nl$, that matches the wavelengths of three lasers, so that the residual Doppler shift is reduced to levels that are comparable to the natural linewidths of the Rydberg states, $\sim$kHz. This scheme will improve the readout substantially and is compatible with the MZI method presented here. The relatively narrow spectral bandwidth of the probe laser transmission window achieved with the MZI allows us to more easily detect the dephasing effects that are important for optimizing the conditions for RF electric field measurements, e.g., transit time broadening and collision broadening.

In the regime of weak RF electric field amplitudes, the signal is sensitive to the probe and coupling laser Rabi frequencies. Fig.~\ref{fig:power} shows the effect of changing the Rabi frequencies of the probe and coupling laser on the change in probe laser transmission on resonance. When the RF electric field amplitude is several hundred $\mathrm{\mu V \, cm^{-1}}$, the probe laser light passing through the vapor cell under the influence of the RF electric field can vary from increased transmission to absorption as the Rabi frequencies of the probe and coupling lasers are changed. The RF electric field induces increased transmission when the Rabi frequencies of the probe and coupling lasers are increased because the EIT window becomes broadened. The broadening of the EIT transmission window and the Doppler averaging can conspire to increase the probe transmission on resonance for weak RF electric field amplitudes \cite{sedlacek2012microwave}. When the EIT conditions are closer to the weak probe regime, the EIT lineshape is narrow and the RF electric field induces more absorption of the probe laser on resonance. The broadened case is the same effect observed and described in Ref.~\cite{sedlacek2012microwave}.

The shape of the curves in Fig.~\ref{fig:power} show that as the RF electric field amplitude is decreased it becomes more difficult to detect changes in the probe laser absorption. This is expected since the RF Rabi frequency is decreasing relative to that of the probe and coupling lasers, as well as other dephasing mechanisms, such as collisions. For weak RF electric field measurements, it is important to have fixed probe and coupling laser Rabi frequencies as well as fixed vapor cell temperature. The atomic density and laser beam size are also important for optimizing the RF electric field measurements.

\begin{figure}[htbp]
\centering
\includegraphics[width=\columnwidth]{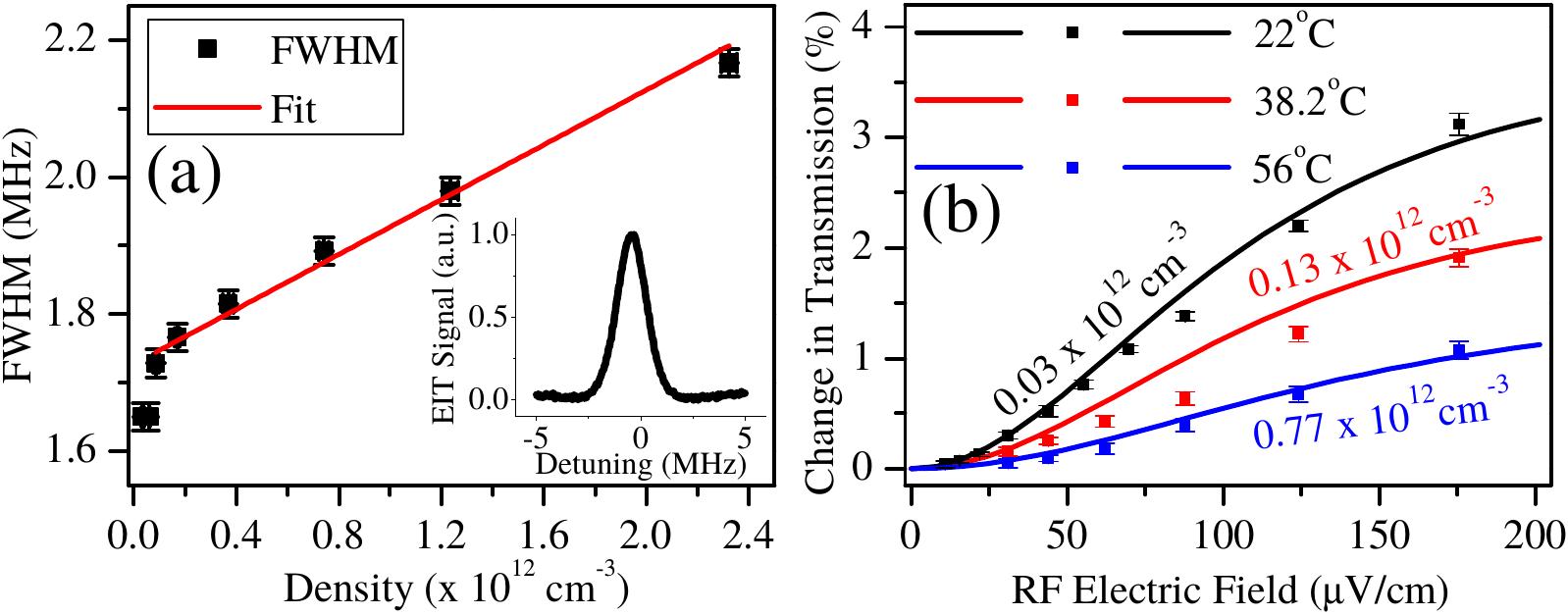}
\caption{The effect of collisional broadening in the measurement of the RF electric field. (a) The FWHM of the EIT probe transmission peak plotted as a function of atomic density. The inset is a typical EIT probe transmission peak at room temperature. The linear dependence of the linewidth on atomic density suggests that the observed broadening is due to collisions. The broadening rate extracted from the linear fit is $2.00\pm 0.13 \times 10^{-13}\,$cm$^3\,$MHz. The corresponding scattering cross-section is $\sigma = 7.2 \times 10^{-12}\,$cm$^2$. $\mathrm{3.1\times10^{10}\,cm^{-3}}$ is the density at room temperature. (b) The change in probe transmission with the coupling and probe laser on resonance induced by the RF electric field for different vapor cell temperatures (densities). For small RF electric field amplitudes, as shown in the figure, collisions can make the measurement less sensitive. $\Omega_p= 2\pi \times 1.3 \pm 0.10\,$MHz and $\Omega_c = 2\pi \times 0.80 \pm 0.02\,$MHz. The coupling beam size is $0.50 \pm 0.005\,$mm. \label{fig:mzi_temp}}
\end{figure}

The effect of changing the Cs atomic density is shown in Fig.~\ref{fig:mzi_temp}. We vary the Cs density in the vapor cell by changing the temperature. The Rydberg atom density is low, so that Rydberg atom-Rydberg atom interactions are negligible. The density of Rydberg atoms in the gas is around $\mathrm{\sim 0.1\%}$ of the ground state atomic density. The primary reason for a low Rydberg atom density is that only a small fraction of the atoms are in the correct velocity class so they meet the EIT condition for the laser beams. The average Rydberg atom interatomic separation is $\sim 20\,\mu$m at room temperature. At these interatomic separations, the interactions between Rydberg atoms are weak and well described by Van der Waals interactions \cite{AAMOP2014}, $<300\,$Hz at room temperature. Even at the highest densities shown in Fig.~\ref{fig:mzi_temp}, the Rydberg-Rydberg interactions are $\sim$kHz. The probe transmission peak broadening shown in Fig.~\ref{fig:mzi_temp}a is due to collisions between Cs Rydberg and ground state atoms \cite{Omont77,Weber83}. The two primary effects that contribute to collisional broadening are elastic and the inelastic scattering of the Rydberg electrons from ground state perturbers. According to Refs.~\cite{Omont77,Weber83}, the FWHM contribution to the line broadening due to these types of elastic collisions is
\begin{equation}
\Gamma_\mathrm{col}^e = 7.18 (\alpha^2 v)^\frac{1}{3} \rho = 1.2 \times 10^{-13}\,\mathrm{cm}^3\,\mathrm{MHz} \times \rho,
\end{equation}
where $\alpha = 402 \,a_0^3$ is the polarizability of ground state Cs, $v$ is the mean velocity of the gas, and $\rho$ is the ground state Cs density in the vapor cell. $v$ changes insignificantly over the temperature range shown in Fig.~\ref{fig:mzi_temp}a. Likewise, the line broadening due to inelastic collisions between the Rydberg electrons and ground state perturbers can be calculated as,
\begin{equation}
\Gamma_\mathrm{col}^i = \frac{8 e^2 a_s^2}{(4 \pi \epsilon_0) h n^*} \rho = \frac{2.2 \times 10^{-12}}{n^*} \,\mathrm{cm}^3\,\mathrm{MHz} \times \rho,
\end{equation}
where $a_s = -16.6\, a_0$ \cite{Thumm01} is the s-wave scattering length for an electron scattering from a ground state Cs atom averaged over the singlet and triplet channels and $n^*$ is the effective quantum number of the Rydberg state. These collisional broadening rates are sufficient to explain our observations, however, at lower $n^*$ there are oscillations in these rates that cannot be explained using only these formulas \cite{gallagher2005rydberg}. These dephasing mechanisms are related to ultra-long range Cs Rydberg molecules \cite{Tallant12,Booth15}. The overall collision rate in this picture is $\Gamma_\mathrm{col} = \Gamma_\mathrm{col}^e + \Gamma_\mathrm{col}^i$. For Cs $\mathrm{52D_{5/2}}$ as the upper state of the EIT system, $\Gamma_\mathrm{col}/\rho= 1.7 \times 10^{-13}\,$cm$^{3}\,$MHz which is in reasonable agreement with the experimentally determined slope of $2.00\pm 0.13 \times 10^{-13}\,$cm$^3\,$MHz shown in Fig.~\ref{fig:mzi_temp}a. Using $\Gamma_\mathrm{col} = \sigma v \rho$, we can report a cross-section for Cs $\mathrm{52D_{5/2}}$ self-broadening of $\sigma = 7.2 \times 10^{-12}\,$cm$^2$. $\sigma$ is similar to other alkali Rydberg- ground state atom cross-sections  that are reported in Ref.~\cite{gallagher2005rydberg}. The discrepancy between the measured and calculated collisional broadening rates could be attributed to the uncertainty in $a_s$ due to uncertainty in the singlet and triplet scattering lengths, the uncertainty in the ground state distribution of atomic hyperfine states due to optical pumping and stray magnetic fields, and the effects of higher order, p-wave, scattering resonances that are prominent for Cs($6S_{1/2}$)-electron scattering \cite{Markson16}.

One interesting consequence of the preceding results is that it is further evidence that superradiance is not playing a significant role in our experiments, consistent with our prior work \cite{dispersive}, despite the fact that $\lambda_\mathrm{RF}$ is much larger than the spacing between Rydberg atoms. The prevalence of Rydberg atom-ground state atom collisions is important in many vapor cell experiments. Perhaps, Rydberg molecule formation, self-broadening and their interplay can partially explain why superradiance is difficult to observe in Rydberg atom vapor cell experiments.

Fig.~\ref{fig:mzi_temp}b shows how the RF electric field measurement depends on the density of atoms in the vapor cell. It becomes more difficult to determine the changes in probe transmission due to the incident RF electric field as collisions dephase the atomic dipoles at rates similar to the RF Rabi frequencies. The slope, which determines the accuracy with which the RF electric field can be measured, decreases as the Cs atomic number density increases. The shape of the curves in Fig.~\ref{fig:mzi_temp}b and Fig.~\ref{fig:power} are similar because a similar effect is happening in both plots. As the RF Rabi frequency decreases, other effects such as collisions begin to overwhelm the effect of the RF electric field, making lineshape changes more difficult to detect. The theory curves in the figures show that the relationship between probe and coupling Rabi frequency and dephasing rates are well described by the density matrix equations.

\begin{figure}[htbp]
\centering
\includegraphics[scale=0.4]{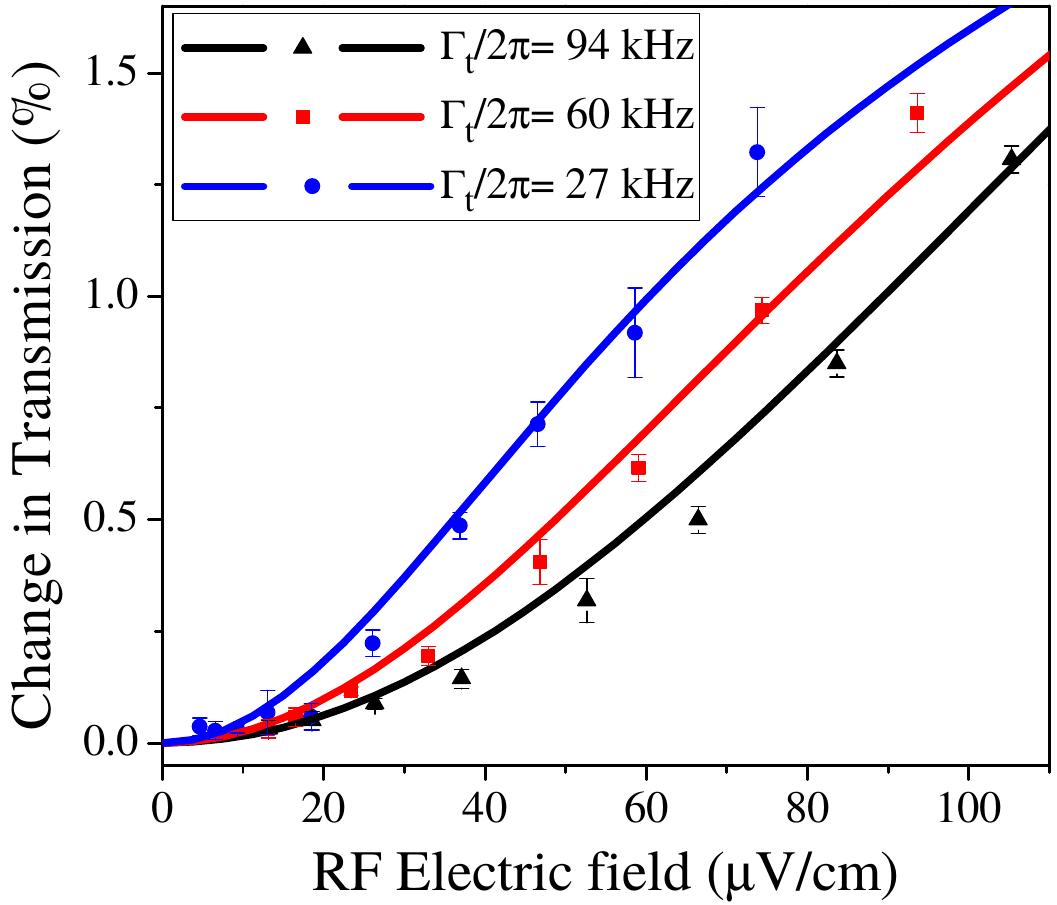}
\caption[short caption]{The effect of transit time broadening in the measurement of the RF electric field. The percentage change in the probe transmission vs. RF electric field amplitude with the probe and coupling lasers on resonance is plotted for three different transit time broadening rates. The transit time broadening rate is varied by changing the coupling laser beam size, the smaller of the two laser beams.  The three coupling laser beam sizes are $\mathrm{0.320\pm 0.005~mm}$, $\mathrm{0.500\pm 0.005~mm}$ and $\mathrm{1.100\pm 0.005~mm}$, corresponding to transit time broadening of $\mathrm{2 \pi \times 94~kHz}$ (black), $\mathrm{2 \pi \times 60~kHz}$ (red) and $\mathrm{2 \pi \times 27~kHz}$ (blue), respectively. The dots are the measurement data and the lines are the numerical results. $\Omega_p = 2 \pi \times 1.7\pm 0.1\,$MHz and $\Omega_c = 2 \pi \times 0.70\pm0.02\,$MHz.\label{mzi_beamsize}}
\end{figure}

In a similar fashion, the laser beam sizes matter because transit time broadening rates can be similar in magnitude to the collision rates and RF Rabi frequencies. The laser beam size also determines the required laser powers needed to achieve a particular Rabi frequency and, for large RF frequencies, the laser beam sizes are limited by the requirement that the vapor cell dimensions be much smaller than $\lambda_\mathrm{RF}$. The transit time broadening, $\mathrm{\Delta \nu}$, is determined by the diameter of the laser beam, $\mathrm{d}$, and the average velocity of the atoms, $v$, $\mathrm{\Delta \nu=\sqrt{2}}v/\mathrm{d}$. Fig.~\ref{mzi_beamsize} shows the effect of the beam size on the RF electric field measurement at room temperature. As the transit time broadening rate increases, the slope of the curve at weak RF electric field amplitudes decreases, making the measurement less sensitive, in a similar fashion to Fig.~\ref{fig:mzi_temp}b and Fig.~\ref{fig:power}. Reducing transit time broadening increases the sensitivity. The experimental results are consistent with the numerical results as shown in Fig.~\ref{mzi_beamsize}.

The weak RF electric field data in Fig.~\ref{fig:mzi_temp}b, Fig.~\ref{fig:power} and Fig.~\ref{mzi_beamsize} allows us to estimate the sensitivity from the measurements presented here as $ \mathrm{\sim 5\mu V cm^{-1} Hz^{-1/2}}$. Each data point corresponds to a $\sim 1\,$Hz bandwidth when averaging, sweep rate, and filtering are taken into consideration. The error bars at small electric fields reflect the sensitivity of the measurements using the MZI. The results are $\sim 6$ times better than our prior results \cite{sedlacek2012microwave}.


The sensitivity achieved with the MZI is around three orders of magnitude worse than the shot noise, or projection noise, limit of the atoms in the vapor cell. The shot noise limited sensitivity in a $1\,$Hz bandwidth of the atoms used for the RF electric field sensor can be calculated as \cite{fan2015atom}
\begin{equation}
\frac{E_{min}}{\sqrt{\mathrm{Hz}}}=\frac{h}{\mu \sqrt{T_{2}N}},
\label{eqn:shotE}
\end{equation}
where $N$ is the effective number of participating Rydberg atoms and $T_{2}$ is the dephasing time. For the room temperature measurements presented in this work, the shot noise limited sensitivity is $\mathrm{\sim 9~nV cm^{-1} Hz^{-1/2}}$. The interaction volume is determined by the $1\,$mm beam diameter and $4\,$cm vapor cell length. For this calculation we took $\mu= 1745~e a_{0}$ and $T_2 = 5\,\mu$s, consistent with $\Gamma_t = 2\pi \times 27\,$kHz, $\Gamma_c = 2\pi \times 6\,$kHz, $\Gamma_m = 2\pi \times 50\,$kHz, and $\Gamma_l= 2\pi \times 70\,$kHz.

The shot noise of the atoms in the vapor cell is not limiting the sensitivity. There are several other sources of noise in the experimental setup. For example, the amplitude-modulation of the coupling laser creates noise that can reduce the sensitivity. Phase noise in the MZI can also degrade the sensitivity. However, a straightforward analysis of the photon shot noise on the photodector implies that improvements in the coupling laser noise and interferometer stability will only be incremental. The SNR of a photon shot noise limited detector is $\sqrt{2 \eta e^2 P_\mathrm{s} \Delta f/ h\nu}$, where $\eta$ is the quantum efficiency, $\Delta f$ is the detection bandwidth, $P_\mathrm{s}$ is the power falling on the detector, and $\nu$ is the frequency of the light \cite{Boyd}. For our detector, with $\sim 10\,\mu$W of probe power falling on the detector in a $1\,$Hz bandwidth, the shot noise limited SNR is $\sim 3 \times 10^6$. The signals shown in Fig.~\ref{fig:mzi_temp}b and Fig.~\ref{mzi_beamsize} are $< 0.1\%$ of the EIT signal while the overall EIT signal without the RF electric field is $\sim 0.1\%$ of the absorption signal. These estimates lead us to conclude that photon shot noise on the photodetector is limiting the sensitivity.

\begin{figure}[htbp]
\centering
\includegraphics[width=4.5in]{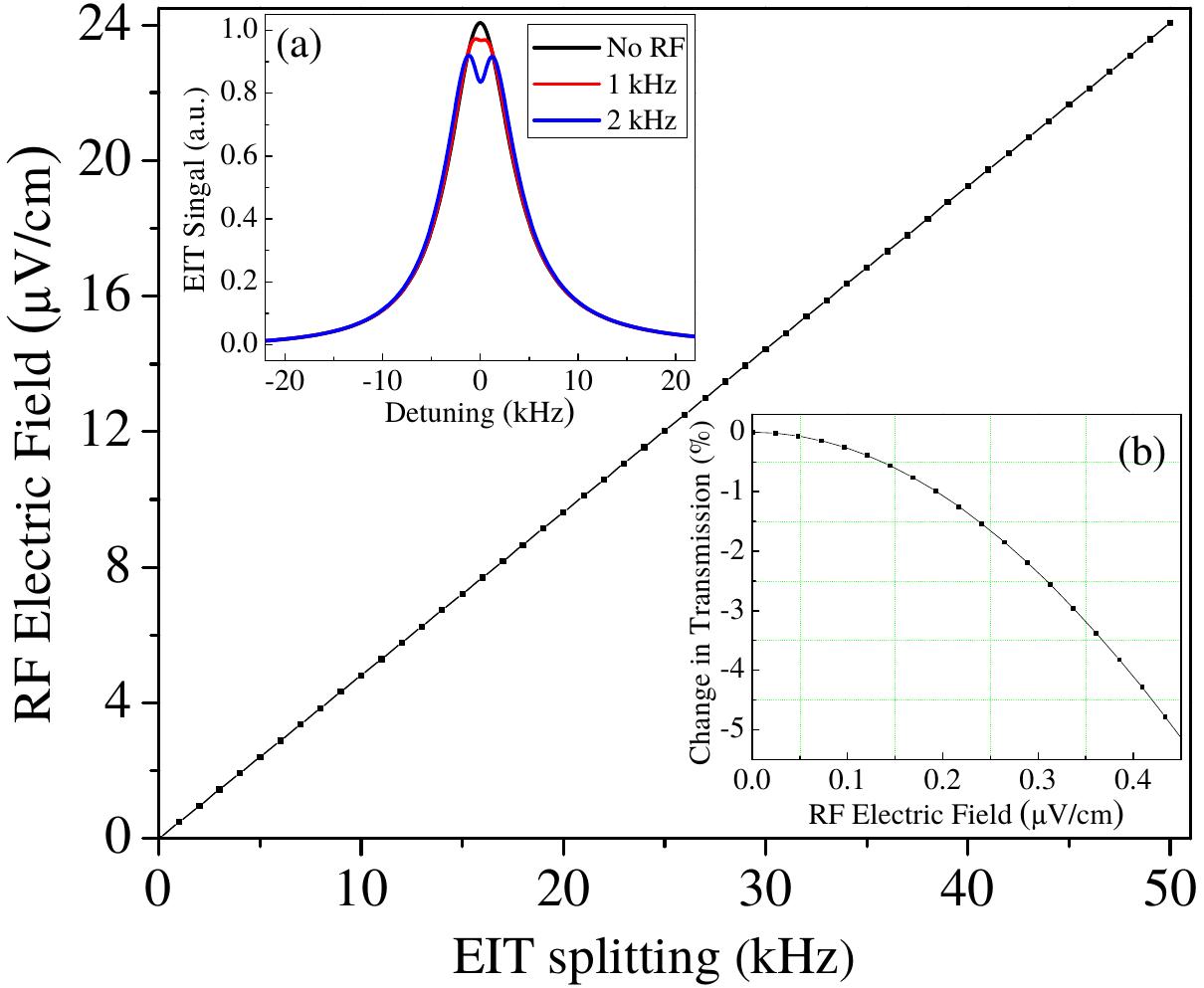}
\caption[short caption]{This figure shows density matrix calculations at room temperature of the three photon detection process, Cs, $6S_{1/2}\leftrightarrow 6P_{1/2} \leftrightarrow 9S_{1/2} \leftrightarrow 53 P_{3/2}$. The main figure shows the RF electric field amplitude as a function of calculated EIT splitting. Inset (a) shows the probe transmission signal for several values of the RF Rabi frequency, $\Omega_p = 2\pi \times 1.8\,$MHz, $\Omega_s = 2\pi \times 1.8\,$MHz, and $\Omega_c = 2\pi \times 50\,$kHz. Here, $p$ denotes the laser tuned close to the Cs $6S_{1/2}\leftrightarrow 6P_{1/2}$ transition, $s$ denotes the laser tuned near the $6P_{1/2} \leftrightarrow 9S_{1/2}$ transition, and $c$ denotes the laser tuned to the $9S_{1/2} \leftrightarrow 53 P_{3/2}$ transition. The RF electric field is tuned to the $52 D_{5/2}\leftrightarrow 53 P_{3/2}$ transition for the sake of comparison. The lasers for $p$ and $s$ are detuned by $500\,$MHz from the intermediate state. The $p$ laser light is detected. The $c$ laser acts as the coupling laser. The calculations include Doppler averaging. The beam size is $5\,$mm. $c$ and the RF electric field were detuned by $5\,$kHz to compensate for asymmetries in the lineshape due to the interplay between Doppler and light shifts. For a photon shot noise limited detection, we predict that we can detect a $\sim 1\%$ change in transmission. This translates to a sensitivity of $\sim 500\,$nV$\,$cm$^{-1}\,$Hz$^{-1/2}$ for a measurement of the peak splitting and an amplitude change detection sensitivity of $\sim 200\,$nV$\,$cm$^{-1}\,$Hz$^{-1/2}$. Inset (b) shows the calculated percent change in transmission that must be determined on resonance to measure different electric fields under the same conditions as inset (a). \label{photon3}}
\end{figure}

The three photon readout we introduced earlier has the potential to improve the sensitivity and accuracy of the method by expanding the AT regime. In the AT regime, the peak splitting is linear to first order in the RF electric field and is easier to determine than a change in peak amplitude. Fig.~\ref{photon3} shows calculations of the three photon readout for the Cs $6S_{1/2}\leftrightarrow 6P_{1/2} \leftrightarrow 9S_{1/2} \leftrightarrow 53P_{3/2}$ system. The first two steps of the excitation are detuned from $6P_{1/2}$ by  $2\pi \times 500\,$MHz so that $6P_{1/2}$ is effectively adiabatically eliminated. The $6S_{1/2}\leftrightarrow 6P_{1/2}$ ($\sim 895\,$nm) light is detected as the probe. The probe Rabi frequency can be large, in order to mitigate photon shot noise, but still have an effectively narrow transmission window. The price is that the power of $\sim 895\,$nm light must increase and the peaks can be become asymmetric due to light shifts and the Doppler effect. The three photon readout can increase the sensitivity by around one order of magnitude. The sensitivity is higher but remains photon shot noise limited. For our example, the atom shot noise limit decreases by $\sim 5$ for the parameters used in Fig.~\ref{photon3} because the dephasing time has increased and the increase in beam size partially compensates the decrease in Rydberg atom number due to the reduced effective Rabi frequencies. The photon shot noise limited SNR also increases with the beam size. Taking into account our current signal levels, the photon shot noise limited SNR, and our calculations of the three photon readout, we conservatively estimate that a $1 \%$ change in the signal is resolvable. This corresponds to a $\sim 1\,$kHz RF electric field Rabi frequency which, for the Cs $52 D_{5/2}\leftrightarrow 53 P_{3/2}$ transition, corresponds to a sensitivity of $\sim 500\,$nV$\,$cm$^{-1}\,$Hz$^{-1/2}$ for a peak splitting measurement and $\sim 200\,$nV$\,$cm$^{-1}\,$Hz$^{-1/2}$ for an amplitude measurement, Fig.~\ref{photon3}. Details of the three photon calculations will be the subject of a forthcoming paper.

The three photon calculations imply that further improvements in the sensitivity needed to achieve the atom shot noise limit require a more sophisticated approach. Photon shot noise is a serious challenge for Rydberg atom-based RF electric field sensing. One approach that can work is to use squeezed light for the probe transition because it can reduce the photon shot noise. Although using squeezed light is challenging, it is worth the effort. Projection noise limited performance would make it possible, for example, to make absolute electric field measurements of thermal background radiation which is fundamentally interesting and would open up new opportunities in precision measurement, for example, studies of blackbody radiation.

\section{Conclusion}

In conclusion, we used a MZI to cancel read-out noise of the probe laser for atom-based RF electric field sensing. We achieved a new sensitivity limit for the absolute measurement of RF electric fields, $\mathrm{\sim 5~\mu V cm^{-1} Hz^{-1/2}}$ which was $\sim 6$ times better than our previous work. We showed the effects of key dephasing mechanisms on the measurements.  Transit time broadening, collision broadening, and power broadening were addressed. Our sensitivity is around three orders of magnitude worse than the shot noise limit of the atomic sensor. By considering the mechanisms for the additional noise, we conclude that photon shot noise on the photodetector is currently a limiting factor. We suggested a three photon method for reading out the RF electric field. This approach can make it possible to use the AT splitting to determine weak RF electric fields. We predict that the three photon readout will increase the sensitivity by an order of magnitude, but will still result in photon shot noise limited performance in many cases. Using squeezed light for the probe laser may lead to better sensitivity. Although we demonstrated our approach with a free space MZI, it is possible to miniaturize the MZI for RF electric field measurement. A chip based, integrated MZI \cite{Ritter_APL2015} or fiber MZI is possible to realize and are likely to be the most useful approaches for RF electric field sensing. Our results also show that it is possible to play off power broadening against sensitivity to increase the signal acquisition rate.

\section{Acknowledgements}

This work was supported by the DARPA Quasar program by a grant through the ARO (60181-PH-DRP). We also acknowledge support from the National Reconnaissance Office. HK acknowledges support from the Carl-Zeiss Foundation.

\newpage


\end{document}